\documentclass[letterpaper, 10 pt, conference]{ieeeconf}  

\IEEEoverridecommandlockouts                              

\usepackage{amsmath} 
\usepackage{amssymb}  
\usepackage{breqn}
\usepackage{svg}
\usepackage{algorithm}
\usepackage{algpseudocode}
\usepackage{mathrsfs}
\usepackage[capitalize]{cleveref}
\usepackage{titlesec}

\title{\LARGE \bf
A Traffic Adapative Physics-informed Learning Control for Energy Savings of Connected and Automated Vehicles
}

\author{Yunli Shao
\thanks{Yunli Shao (yunli.shao@uga.edu) is with the School of Environmental, Civil, Agricultural and Mechanical Engineering, University of Georgia, Athens, 30602}
}

\begin{document}

\maketitle
\thispagestyle{empty}
\pagestyle{empty}

\setlength{\abovedisplayskip}{4pt}
\setlength{\belowdisplayskip}{4pt}

\begin{abstract}
Model predictive control has emerged as an effective approach for real-time optimal control of connected and automated vehicles. However, nonlinear dynamics of vehicle and traffic systems make accurate modeling and real-time optimization challenging. Learning-based control offer a promising alternative, as they adapt to environment without requiring an explicit model. For learning control framework, an augmented state space system design is necessary since optimal control depends on both the ego vehicle’s state and predicted states of other vehicles. This work develops a traffic adaptive augmented state space system that allows the control strategy to intelligently adapt to varying traffic conditions. This design ensures that while different vehicle trajectories alter initial conditions, the system dynamics remain independent of specific trajectories. Additionally, a physics-informed learning control framework is presented that combines value function from Bellman’s equation with derivative of value functions from Pontryagin’s Maximum Principle into a unified loss function. This method aims to reduce required training data and time while enhancing robustness and efficiency. The proposed control framework is applied to car-following scenarios in real-world data calibrated simulation environments. The results show that this learning control approach alleviates real-time computational requirements while achieving car-following behaviors comparable to model-based methods, resulting in 9\% energy savings in scenarios not previously seen in training dataset.
\end{abstract}

\section{Introduction}
Over the past decades, connected and automated vehicles (CAVs) have attracted significant attention for their potential to enhance energy efficiency, mobility, and safety in automotive and transportation systems. Connectivity allows effective prediction of future traffic conditions, enabling the design of optimal control strategies that can be executed through vehicle automation. Traditionally, model-based control strategies are applied for CAV design \cite{vahidi2018energy,shaovtm}. These strategies often rely on gradient-based numerical optimization solvers to find optimal solutions in real time. With advancements in onboard computational power, model predictive control (MPC) has emerged as an effective approach for real-time implementation. However, the effectiveness of MPC depends on the availability of accurate models. Given the highly nonlinear dynamics involved in vehicle and traffic systems, modeling these complexities poses challenges and can significantly increase computational demands \cite{xu2020preview, stern2018dissipation}. A simplified model may reduce computational burdens but at the cost of performance and optimality. Additionally, physical systems often include both discrete and continuous states, such as discrete gear positions in a vehicle \cite{shao2020vehicle}, which adds complexity for gradient-based search algorithms.

On the other hand, learning-based algorithms have gained popularity because they learn optimal control strategies directly by interacting with the environment. Learning control (LC) methods include model-based approaches such as approximate dynamic programming (ADP) \cite{powell2007approximate,kargar2022optimal}, where the value function in the Bellman equation is learned based on knowledge of the system dynamics, and model-free approaches like reinforcement learning (RL) \cite{sutton2018reinforcement,dong2020deep}, which learns both the value function and system on the fly. RL is promising because it can adapt to the environment without needing an explicit model at first place. However, its application in transportation systems remains challenging due to inherent complexity of traffic dynamics \cite{haydari2020deep}. Applying RL directly to real-world traffic is often impractical, requiring use of simulated environments or system models to facilitate the initial learning phase of optimal control strategies.

\begin{figure}[h!]
  \centering
  \includesvg[inkscapelatex=false,width=0.44\textwidth]{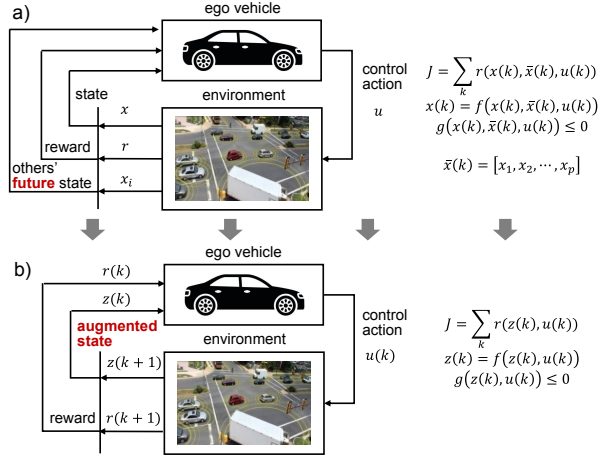}
  \caption{Original Optimal Control Problem Versus Learning Control}
  \label{fig:highlevel}
\end{figure}

From the ego vehicle's perspective, the original optimal control problem (OCP) is illustrated in \cref{fig:highlevel}a. The optimal control action depends not only on the ego vehicle’s state but also on the traffic scenarios, which require information about the states of other vehicles. Since optimal control problems are formulated for a look-ahead horizon, both the current and predicted future states of other vehicles are necessary. As shown in \cref{fig:highlevel}b, to implement a learning control algorithm, the original OCP must be reformulated with augmented states that include both the ego vehicle's state and those of other vehicles. While it might seem feasible to include only the current states of other vehicles, this approach essentially relies on the learning algorithm to predict other vehicles’ future behaviors implicitly. Unlike typical autonomous driving frameworks, which separate prediction and planning \& control into distinct modules, this approach merges prediction and control into a single system. This can introduce additional layers of complexity, requiring more extensive training data, larger computational resources, and potentially making the system challenging to explain and troubleshoot. Furthermore, the interactions between nearby vehicles also affect the ego vehicle’s optimal control decisions, making it essential to consider the states of all relevant vehicles. Predicting other vehicles’ behaviors is a complex research area on its own \cite{karle2022scenario,mahjourian2022occupancy}, introducing additional challenges when integrated into the learning control framework. 

This work considers an autonomous driving pipeline with separate prediction and planning \& control modules, similar to that in previous work \cite{shaovtm}. The prediction module generates future trajectories or states of other vehicles, which serve as inputs to the optimal control module. Since the optimal control strategy depends on these predicted future states, an explicit transcription of these states is required, and an augmented state space system must be designed accordingly.

\vspace{-3mm}
\begin{figure}[h!]
  \centering
  \includesvg[inkscapelatex=false,width=0.45\textwidth]{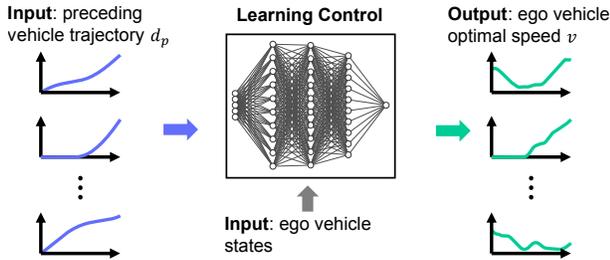}
  \caption{Design Objective of Traffic Adaptive Learning Control}
    \label{fig:trafficdaptive}
\end{figure}

The main contributions of this work are twofold. First, it develops a traffic adaptive augmented state space system that incorporates the predicted future states of preceding vehicles. The proposed formulation ensures that different vehicle trajectories only change the initial conditions while keeping the system dynamics independent of the specific trajectory. As illustrated in \cref{fig:trafficdaptive}, this design enables the learned control strategy to intelligently adapt to varying traffic conditions: even if the ego vehicle’s state remains the same, its control strategy will adjust based on the predicted trajectories of other vehicles, enabling adaptive behavior in dynamic traffic environments. Second, a physics-informed learning control (PILC) framework is introduced that combines the value function from the Bellman equation with the derivative of value functions from Pontryagin’s Maximum Principle (PMP) into a unified loss function. Inspired by recent advances in physics-informed neural networks (PINNs) \cite{raissi2019physics}, this approach has the potential to reduce the required amount of training data and time, thus enhancing robustness and efficiency in the learning control process.

In the remainder of this paper, both the PILC framework and the augmented system design are presented in detail. The proposed traffic adaptive PILC is then applied to car-following scenarios \cite{shaoits}, where the ego vehicle follows a preceding vehicle based on its predicted future states. The objective for the ego vehicle is to optimize its following distance to improve energy efficiency. While this work focuses on the car-following scenario, the proposed framework is flexible and extendable to more complex driving scenarios or other control objectives in future studies.

\section{Physics-informed learning control framework}

\subsection{Generic Learning Control Framework}
Consider a discrete system with index $k=0,1,2,\cdots, N$. and assume the discrete time step is $\delta t$:
 \begin{align}
     x_{k+1}=f(x_k,u_k)
 \end{align}

For simplicity, assume discount factor is $\gamma=1$. Suppose the objective is to design control $u$ to maximize 
\begin{align}
    \max_{u_k} \quad & J = \sum_{k=0}^{N-1} r(x_k,u_k)
\end{align}

The optimal control $u^*$ satisfies Bellman equation \cite{sutton2018reinforcement}
\begin{align}
\label{eq:bellman}
V_k(x_k) = r(x_k, u_k^*) + V_{k+1}(f(x_k, u_k^*)) 
\end{align}
where $V_k(x_k)$ is the value function (or known as cost-to-go):
\begin{align}
V_k(x_k)=\sum_{i=k}^{N-1} r(x_i,u_i)
\end{align}
 
The task of typical learning control algorithms (e.g., ADP, RL) is to utilize approximation functions such as neural networks (NNs) to estimate value function $V(x_k,\phi)$, also known as the critic \cite{sutton2018reinforcement}, with $\phi$ as weights and biases of the NN. Often an infinite horizon optimal control problem ($N\rightarrow\infty$) is considered and subscript of value function is dropped. Another NN is designed to approximate the optimal control strategy as a function of state, known as actor. That is $u^*=\pi (x_k, w)$ where $w$ are weights and biases.

In general, the critic is learnt to minimize the following loss function to satisfy the Bellman equation \eqref{eq:bellman}:
\begin{align}
    J_{\phi}=\left\Vert V(x_k,\phi)-(r(x_k,u_k) + V(x_{k+1},\phi) )\right\Vert^2
\end{align}

In practice, there are many variants \cite{dong2020deep}, including versions where $Q(x_k,u_k)$ is learned \cite{dong2020deep,sutton2018reinforcement}. Nevertheless, the general principle of learning control holds and it is the intention of this paper to formulate the generic learning control framework, as summarized in \cref{alg:genericLC}, which can be extended for other specific implementation.

\begin{algorithm}
\caption{Generic Learning Control Algorithm}
\label{alg:genericLC}
\begin{algorithmic}
\State \textbf{Input:} Initialize $\phi,w$
\State $N \gets n$
\For{each episode}
    \For{each step in episode}
        \State run policy $u=\pi (x_k, w)$ and observe $x_{k+1}$ and $r_k$
        \State update $\phi$ to minimize $J_{\phi}$
        \State update $w$: maximize $r(x_k,\pi(x_k,w)) + V(x_{k+1},\phi)$
    \EndFor
\EndFor
\end{algorithmic}
\end{algorithm}

\subsection{Physics-informed Learning Control}
As can be seen above, the learning control algorithm are primarily based on the Bellman equation \eqref{eq:bellman}. If we take derivative on both side of \eqref{eq:bellman} with respect to $x_k$, then
\begin{align}
\label{eq:pmplambda}
\frac{\partial{V_k(x_k)}}{\partial x_k} = \frac{\partial{r(x_k, u_k^*)}}{\partial x_k} + \frac{\partial{V_{k+1}(x_{k+1})}^T}{\partial x_{k+1}}\frac{\partial{f(x_k, u_k^*)}}{\partial x_k}
\end{align}

Define costates $\lambda_k=\frac{\partial{V_k(x_k)}}{\partial x_k}$, then \eqref{eq:pmplambda} essentially follows the necessary conditions from discrete PMP \cite{jin2020pontryagin}:
\begin{align}
\label{eq:pmp}
    \lambda_k &= \frac{\partial H_k}{\partial x_k}=\frac{\partial r}{\partial x_k}+\lambda_{k+1} \frac{\partial f}{\partial x_k}, \\
    u^*&=\arg \min_u H_k
\end{align}
where $H_k$ is the Hamiltonian defined as
\begin{align}
    H_k = r(x_k,u_k) + \lambda_{k+1}^T f(x_k,u_k)
\end{align}

\eqref{eq:pmplambda} essentially gives another set of conditions that the value function should satisfy in forms of differential equations. It is intuitive to add this condition to the learning control algorithm as well. This can be achieved inspired by recent development of the PINN \cite{raissi2019physics}, where solutions to partial differential equations (PDEs) are learned through training a neural network with loss function of both the boundary conditions, in ordinary form, and the PDE, in differential form. The physics-informed loss function is:
\begin{align}
    J_p=\left\Vert \frac{\partial{V_k(x_k)}}{\partial x_k} - (\frac{\partial{r}}{\partial x_k} + \frac{\partial{V_{k+1}(x_{k+1})}^T}{\partial x_{k+1}}\frac{\partial{f}}{\partial x_k}) \right\Vert ^2
\end{align}

Differentials of value function can be obtained by auto-differentiation. Combining loss functions $J_p$, $J_\phi$, a physics-informed learning control framework can be developed: 
\begin{algorithm}
\caption{Physics-informed Learning Control (PILC)}
\label{alg:PILC}
\begin{algorithmic}
\State \textbf{Input:} Initialize $\phi,w$
\State $N \gets n$
\For{each episode}
    \For{each step in episode}
        \State run policy $u=\pi (x_k, w)$ and observe $x_{k+1}$ and $r_k$
        \State {\color{blue} update $\phi$ to minimize $J_{\phi} + w_p J_p$}
        \State update $w$: maximize $r(x_k,\pi(x_k,w)) + V(x_{k+1},\phi)$
    \EndFor
\EndFor
\end{algorithmic}
\end{algorithm}

The above PILC only modifies critic training. If letting the first order derivative of $H_k$ w.r.t. $u$ vanish to satisfy \eqref{eq:pmp}:
\begin{align}
    \frac{\partial H_k}{\partial u_k}=\frac{\partial r}{\partial u_k}+\lambda_{k+1} \frac{\partial f}{\partial u_k}=0
\end{align}

This additional differential equation for optimal control $u^*$ can also be included as the physics-informed loss function for actor update, which will be part of the future work.

Comparing to the learning control in \cref{alg:genericLC}, now we require information of the system state equation $f$, in other words, a system model is required. The accuracy of this model will affect the performance of this learning control algorithm. This model can be learned simultaneously with the control strategy using methods such as model-based RL, which will be part of the future exploration. 
\section{Traffic Adaptive Learning Control using An Augmented State Space System}
In this work, a car-following scenario \cite{shaoits} is studied where the ego vehicle adjusts its following distance intelligently for energy savings based on predicted future trajectory of preceding vehicle. As illustrated in \cref{fig:highlevel}, one single sets of states are needed for the learning control framework. However, the original optimal control problem depends on both the states $x$ of ego vehicle as well as future states of preceding vehicles, such as future locations of preceding vehicles $d_p(k), k=0,1,\cdots,N$. In addition, from the Bellman equation \eqref{eq:bellman}, the value function also depends on the time instance $k$ \cite{heydari2014fixed, kargar2022optimal}. The objective of augmented system design is to incorporate the ego vehicle states $x$, transcribed future states of preceding vehicles (\cref{sec:dptranscribe}), and time index $k$ into a unified set of states $z$. Then, the typical learning control framework can be applied to the augmented system. 
	
\subsection{Original Optimal Control Problem}
The PILC is implemented in a receding horizon control (RHC) fashion. To establish the foundation of a traffic adaptive PILC for CAVs, this work did not consider detailed powertrain dynamics and adopts a point mass vehicle model yet. It is part of the future work to explore incorporation of more detailed powertrain states and complex dynamic models. This car-following optimal control problem is \cite{shaoits}:
\begin{equation}
\begin{aligned}
\max_{u(k)} \quad & J = \psi \cdot (x(k)-x_f)^2 + \sum_{k=0}^{N-1} r(x(k),u(k))\\
\textrm{s.t.} \quad & x(k+1)=f(x(k),\bar{x} (k),u(k))\\
  &  g(x(k)) \leq 0  \\
  &  x(0) = x_0
\end{aligned}
\end{equation}

The states $x(k)$ include $d_f$, car-following distance between ego and preceding vehicles, and speed of ego vehicle $v$. The control input $u$ is the acceleration of the ego vehicle. The car following distance at $k+1$ can be written as:
\begin{equation}
\begin{aligned}
\label{eq:stateeq}
& d_f(k+1) = d_p(k+1)-d(k+1) \\
         & = d_p(k+1)-d_p(k)+d_p(k)-(d(k)+\delta t\cdot v(k)) \\
         & = d_p(k+1)-d_p(k)+d_f(k)-\delta t\cdot v(k)
\end{aligned}
\end{equation}

The ego vehicle speed follows the equation:
\begin{align}
\label{eq:stateeq_v}
v(k+1) & = v(k)+\delta t\cdot a(k) 
\end{align}

The reward (or objective function) includes two parts, power consumption of the vehicle, and the square of acceleration with weighting factor $w_1$:
\begin{align}
\label{eq:r}
r(x(k),u(k))=-(P_{veh}(x(k),u(k))+w_1\cdot u^2(k))
\end{align}

The vehicle power consumption is modeled using the longitudinal dynamics of the vehicle \cite{shaovtm}:
\begin{equation}
\begin{aligned}
P_{veh}(k) & = (\mu mg cos(\theta(k)) + mg sin(\theta(k)) \\
                       & + 1/2 C_d \rho A v^2(k) + m\cdot u(k))\cdot v(k)
\end{aligned}
\end{equation}

For simplicity, only car-following distance constraints are considered, which enforce the ego vehicle to keep a safe headway $h_s$ with minimum distance $d_{min}$, and maintain satisfactory traffic throughput with maximum distance $d_{max}$:
\begin{align}
    g(x(k))= \begin{bmatrix}
        d_f(k) - d_{max} \\
        -d_f(k) + (d_{min}+h_s\cdot v(k))
        \end{bmatrix}
        \leq 0
\end{align}

These car-following constraints can be augmented as part of reward using sigmoid function $SIG$:
\begin{equation}
   \begin{aligned}
   \label{eq:rg}
    & r_g(k) = -w_2 \cdot [SIG(d_f(k)-d_{max}) \cdot (d_f(k) - d_{max})^2 \\
    &    +SIG(d_f(k) - d_{min}-h_s\cdot v(k)) \cdot \\
    & (d_f(k) - d_{min}-h_s\cdot v(k))^2]
\end{aligned} 
\end{equation}

The terminal constraints are needed otherwise the vehicle will tend to glide towards the end to save energy, which could bring in negative impacts to the next control update cycle. The following terminal conditions are enforced to maintain a terminal headway $h_t$ and keep the same speed with the preceding vehicle:
\begin{align}
\label{eq:terminal}
x_f = [d_{min} + h_t\cdot v_p(N), v_p(N)]^T
\end{align}

Initial states $x_0$ are final states from previous RHC step. 


\subsection{Transcribe Predicted Location of Preceding Vehicle}
\label{sec:dptranscribe}
\begin{figure}[h!]
  \centering
  \includesvg[inkscapelatex=false,width=0.5\textwidth]{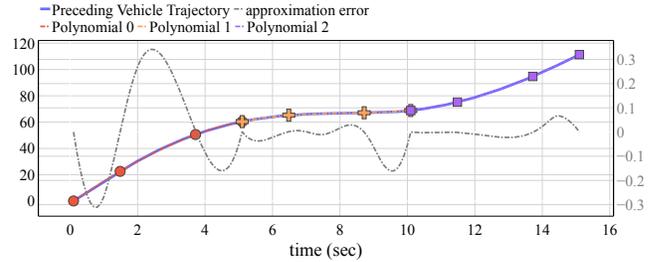}
  \caption{Typical $d_p$ Trajectory and Transcribe Error}
    \label{fig:polyfit}
\end{figure}

At each RHC update horizon, a predicted preceding vehicle trajectory of the next $N\cdot\delta t$ seconds are provided as input to the optimal controller, which consists of $N$ data points. One immediate thought would be directly input all these $N$ data points into the learning control framework. However, $N$ could equal to a few hundreds and such approach could results in a large state space and bring in complexities for learning the value function effectively. Inspired by the pseudospectral method from optimal control \cite{darby2011hp}, this work transcribes trajectory of $d_p(k)$ using a series of polynomials. Each trajectory is divided into several intervals, and a low order polynomial is used to approximate the trajectory of each interval, as illustrated in \cref{fig:polyfit}. 
Other potential methods such as using an encoder NN \cite{abdelraouf2021utilizing} will be part of future work. As can be seen in \cref{fig:polyfit}, with intervals of $5~sec$ and polynomial of order $3$, transcription of typical $d_p(k)$ can achieve accuracy of $\pm0.3~m$. Here $d_p(k)$ is selected since the state equations \eqref{eq:stateeq} only depends on $d_p(k)$. Also, $d_p(k)$ is relatively smoother than $v_p(k)$ and can reduce the required number of intervals and augmented states.

Suppose the time instance $k$ is augmented as another state:
\begin{align}
\label{eq:stateeq_s}
s(k+1) = s(k)+1
\end{align}

Then, the $d_p(k)$ can be expressed as the following: 
\begin{align}
d_p(k) = \sum_{i=0}^{N_c} [\sigma(k-i\cdot T_{c}) \cdot \sum_{j=0}^{N_p}(p_j^i \delta t^j \cdot (s(k)+1)^j)]
\end{align}
\begin{align}
\sigma(k-i\cdot T_{c}) = 
    \begin{cases}
    0, k < i\cdot T_c~ or ~ k > (i+1)\cdot T_c\\
    1, i\cdot T_c \leq k \leq (i+1)\cdot T_c\\
    \end{cases}
\end{align}
where $d_p$ is transcribed using $N_c$ intervals, each of time duration $T_c$, with polynomial of order $N_p$. At one time instance $t$, only one of the polynomial will be \textit{activated} determined by $\sigma(\cdot)$. Denote the coefficients of the polynomial from interval $i$ as $p_j^i, j=0,\cdots, N_c$. The objective is to keep the system dynamics independent of the specific trajectory $d_p$, but in its current form, $d_p$ depends on the coefficients of the polynomials. The following augments states can be further defined for $j>0$:
\begin{align}
z_j^i(k) = p_j^i \delta t^j \cdot (s(k)+1)^j
\end{align}

These states can be written in standard state space format:
\begin{equation}
\begin{aligned}
\label{stateeq_z}
z_j^i(k+1) & = p_j^i \delta t^j \cdot (k+2)^j \\
           & = z_j^i(k) \cdot \frac{(s(k+1)+1)^j}{(s(k)+1)^j} \\
           & = z_j^i(k) \cdot \frac{(s(k)+2)^j}{(s(k)+1)^j} 
\end{aligned}
\end{equation}

All $z_0^i(k)=p_0^i$ are constant and not preferred to be designated as states. The following variable is defined:
\begin{equation}
\begin{aligned}
\label{stateeq_q}
q_i(k+1) & = z_1^i(k+1) + p_0^i \\
           & = p_1^i \delta t \cdot (k+2) + p_0^i \\
           & = z_1^i(k+1)-z_1^i(k)+p_0^i+z_1^i(k) \\
           & = q_i(k) + z_1^i(k) \cdot (\frac{(s(k)+2)}{(s(k)+1)}-1)
\end{aligned}    
\end{equation}

Therefore, we can write $d_p$ in the forms of states, without explicitly depending on polynomial parameters:
\begin{align}
\label{stateeq_dp_trans}
d_p(k) = \sum_{i=0}^{N_c} [\sigma(k-i\cdot T_{c}) \cdot (\sum_{j=1}^{N_p} z_j^i(k) + q_i(k))]
\end{align}

\subsection{Transcribe Terminal Constraints}
The terminal constraints \eqref{eq:terminal} depends on $v_p(N)$, which can be estimated using the transcribed state $d_p$:
\begin{align}
v_p(N) = \frac{d_p(N)-d_p(N-1)}{\delta t}
\end{align}

Therefore,
\begin{align}
x_f = \begin{bmatrix}
    d_{min} + h_t\cdot  \frac{d_p(N)-d_p(N-1)}{\delta t} \\
    \frac{d_p(N)-d_p(N-1)}{\delta t}
    \end{bmatrix}
\end{align}
where $d_p(N)$ and $d_p(N-1)$ can be replaced by states $z_j^i$ and $q_i$ using relationship \eqref{stateeq_z}\eqref{stateeq_q}. These terminal constraints could be formulated as part of the running reward using ReLU function and time instance state $s(k)$:
\begin{equation}
\begin{aligned}
\label{eq:rpsi}
&r_\psi(k) = -ReLU(k-N) \cdot  \\
   & (\psi_1\cdot(d_f(k) - (d_{min}+h_t \cdot \frac{d_p(k)-d_p(k-1)}{\delta t}))^2 \\
   & +\psi_2 \cdot(v(k) - \frac{d_p(k)-d_p(k-1)}{\delta t})^2 )
\end{aligned}
\end{equation}



\subsection{Augmented State Space System}
The previous formulations are for general scenarios. In this paper, suppose the RHC prediction horizon is $15~sec$ and  $d_p$ is divided into three intervals ($N_c=2$) using a polynomial of order 3 ($N_p=3$), then the augmented states are:
\begin{equation}
\begin{aligned}
\label{eq:augstate}
    z(k) = [d_f(k), v(k), s(k), z_3^0(k), z_2^0(k), z_1^0(k), q_0(k), \\
    z_3^1(k), z_2^1(k), z_1^1(k), q_1(k), z_3^2(k), z_2^2(k), z_1^2(k), q_2(k)]^T
\end{aligned}    
\end{equation}

The state space dynamics of $d_f$ can be obtained by plugging \eqref{stateeq_dp_trans} into \eqref{eq:stateeq}, dynamics of $v$ and $s$ follows \eqref{eq:stateeq_v} and \eqref{eq:stateeq_s}, and $z_j^i$ and $q_i$ follows \eqref{stateeq_z} and \eqref{stateeq_q}. Therefore, the proposed formulation ensures that different vehicle trajectories only change the initial conditions while keeping the system dynamics independent
of the specific trajectory. This way, the learned control strategy will adjust based on the predicted trajectories of preceding vehicles even if the ego vehicle’s state remains the same, enabling adaptive behavior in dynamic traffic environments. At each update horizon, relative position of $d_p$ is used that always starts at $0.1~meter$ and the time instance always starts with $\delta t$ seconds (this is to avoid zero on denominators in \eqref{stateeq_z}). Therefore, the ranges of time instance states $s(k)$ and $z_j^i$ and $q_i$ will be kept within a bounded range during the learning process to improve the algorithm's robustness. The initial values are:
\begin{equation}
    \begin{aligned}
        d_p(0)=0, s(0)=0, v(0)=v_0 \\
        z_j^i(0) = p_j^i\delta t^j, q^i(0) = p_1^i\delta t+p_0^i
    \end{aligned}
\end{equation}
where initial speed $v_0$ is the same as last value from implementing the previous RHC cycle. The augmented reward function is the combination of \eqref{eq:r}\eqref{eq:rg}\eqref{eq:rpsi}:
\begin{align}
\label{eq:augr}
    r_{aug}(k) = \sum_{k=0}^{N-1} r(x(k),u(k)) + r_\psi(k) + r_g(k)
\end{align}

With state dynamics \eqref{eq:augstate} and reward \eqref{eq:augr}, the PILC framework in \cref{alg:PILC} can be applied to learn the optimal control strategy. 
\section{Simulation Results}

\subsection{Training Scenarios}

Training data are generated from a real-world data calibrated VISSIM traffic microsimulation network that represents a corridor at Shallowford Road in Chattanooga, Tennessee. The entire simulation network includes a total of 8 intersections. As a first step to validate the overall PILC framework, data from 200 vehicles around one intersection are used to train both the actor and critic networks. The training lasts $10000$ episodes and each episode is fixed at $15~sec$ that equals to the RHC prediction horizon. The PILC is based on Soft Actor Critic (SAC) \cite{haarnoja2018soft} and the PILC utilizes the relationship $V(x_t)=\mathbb{E}_{u_t\sim\pi}[Q(x_t,u_t)]$ to estimate the gradients in \cref{alg:PILC} utilizing partial derivatives of $Q$. Both the critic and actor network utilized multilayer perceptron (MLP) networks with 3 hidden layers and each of 256 neurons. 


\subsection{Results}
\begin{figure}[h!]
  \centering
  \includesvg[inkscapelatex=false,width=0.5\textwidth]{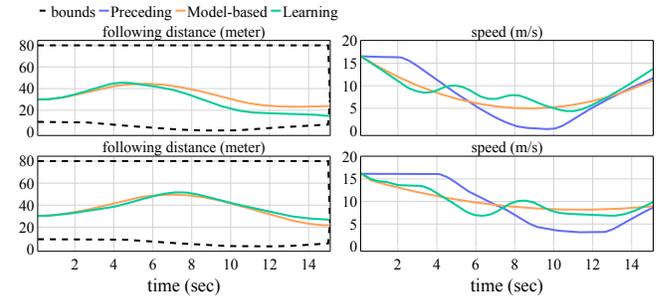}
  \caption{Comparison of MPC versus PILC}
    \label{fig:comparempc}
\end{figure}

The trained PILC is first compared with the model based MPC design \cite{shaoits} for $15~sec$ prediction horizon. \cref{fig:comparempc} shows the comparison of two typical profiles. Overall, the two control strategies show comparable car-following distance behaviors. The speed profile from PILC has more ups and downs, but these variations ``balanced out'' which results in similar car-following behaviors with MPC. The variations of the speed profiles could result from the fact that vehicle power demand is directly penalized in this paper. There lacks penalty for negative power demand or frequent switching sign of power demand and the learning control seeks to fully leverage the opportunities in negative power to reduce the total power demand. Nevertheless, since the PILC is implemented in a RHC framework, only the first few optimal control strategies will be implemented before receving an updated control command. The optimal speed profiles of model-based and learning control are close to each other for the first 2 seconds. So in general, the PILC will result in similar performance with the model-based design with RHC as long as update frequency of less than 2 seconds. Note that one of the benefits of PILC is that it offloads the computation burdens to the training stages. During online implementation, the inference time of the learned controller is much efficient compared to real-time solving a nonlinear programming problem as in model-based MPC. A much faster update horizon could be implemented using PILC to improve the frequency of feedback from RHC to more effectively compensate for prediction errors. Also, PILC has the potential to tackle systems with more complex and accurate dynamic models, such as learned models based on neural networks which can be challenging to solve by model-based MPC. 

\begin{figure}[h!]
  \centering
  \includesvg[inkscapelatex=false,width=0.5\textwidth]{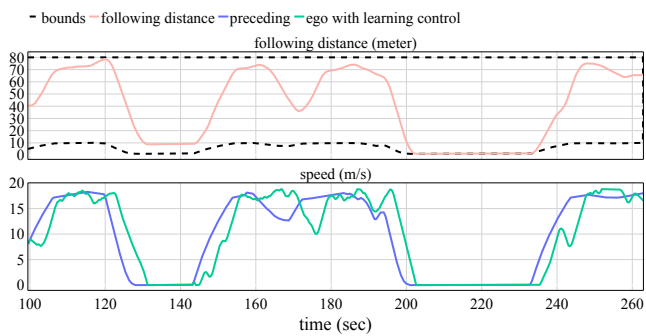}
  \caption{Optimal Control Results of A Selected Vehicle Using PILC}
    \label{fig:fulltraj}
\end{figure}

After training, the PILC was also evaluated for vehicles traveling across the entire corridor, which were clearly not seen during the training stage. It is assumed that the RHC prediction horizon is always 15 seconds and the PILC control is re-calculated every 1 seconds to be consistent with the model-based design MPC. 
\cref{fig:fulltraj} shows the car-following and optimized vehicle speed profiles for a selected vehicle. Overall, the net power demand for the preceding vehicle is 0.197kWh and it is 0.179kWh for PILC with energy savings of 9.1\%. The PILC vehicle reduced power demand during both acceleration and deceleration to reduce energy consumption, and fully utilizes the freedom within the car-following distance bounds. The ups and downs in the speed profiles corresponds to a ``pulse-and-glide'' pattern, which is shown to provide energy benefits \cite{li2015effect}. Such behaviors could cause uncomfortable feelings and an alternative reward function could further improve the smoothness, which will be part of the future work.

For simplicity, it is assumed that the ego vehicle has perfect knowledge of the preceding vehicle trajectory. It is part of the future work to integrate the PILC with a traffic prediction algorithm, such as previous work \cite{shaovtm}. Also, the PILC indeed gives a reference speed profile for the ego vehicle and eventually another lower level vehicle speed tracking controller should be integrated \cite{wang2020ultra}. This is also out of the scope of this work.    

\section{Conclusion}
In this work, we first introduce a physics-informed learning control framework, followed by the design of a traffic adaptive augmented system. The goal is to enhance robustness and efficiency in the learning control process, allowing the learned control strategy to intelligently adapt to varying traffic conditions. This methodology is applied to a car-following scenario focused on energy savings, validated with real-world data calibrated simulation scenarios. Evaluations demonstrate that the resulting control approach achieves car-following behaviors comparable to model-based methods while alleviating real-time computational demands. Moreover, the proposed learning control shows robustness, effectively addressing unseen scenarios with a 9\% energy savings benefit. This work serves as an initial exploration of the physics-informed learning control framework, opening several future research directions. Future work will include investigating different reward functions and transcription methods, applying the framework to more complex and accurate dynamic models, conducting larger-scale validations, incorporating additional physics-informed designs for actor network learning, integrating model-based reinforcement learning to learn dynamic models on the fly, and performing experimental validations.

\bibliographystyle{IEEEtran}
\bibliography{reference}

\end{document}